\documentclass[12pt,onecolumn]{article}

\usepackage{arxiv}
\usepackage{amsmath}
\usepackage{graphics}
\usepackage[utf8]{inputenc} 
\usepackage[T1]{fontenc}    
\usepackage{url}            
\usepackage{booktabs}       
\usepackage{amsfonts}       
\usepackage{nicefrac}       
\usepackage{microtype}      
\usepackage{lipsum}

\usepackage{amsfonts}
\usepackage{amssymb}
\usepackage{bbold}
\usepackage{booktabs}
\usepackage{ulem}
\usepackage{xr}
\usepackage[justification=centering ]{subcaption}
\usepackage{graphicx}
\usepackage{amsmath}
\usepackage[usenames,dvipsnames]{xcolor}
\usepackage{multirow}
\usepackage{lscape}
\usepackage[sorting = none, backend = bibtex, style=phys]{biblatex}
\usepackage{amsfonts}
\usepackage{amssymb}
\usepackage{bbold}
\usepackage{booktabs}
\usepackage{ulem}
\usepackage{xr}
\usepackage[justification=centering ]{subcaption}
\usepackage{tikz}
\usetikzlibrary{shapes}
\usepackage{soul}

\usepackage{nicefrac}
\usepackage{xcolor}
\usepackage{bm}
\usepackage{multirow}
\usepackage{amsfonts}
\usepackage{amssymb}
\usepackage{bbold}
\usepackage{hyperref}
\usepackage{capt-of}
\usepackage{float}
\usepackage{lipsum}
\usepackage{soul}

\usepackage{color}
\usetikzlibrary{shapes}

\usepackage{color}
\usepackage{tikz}
\usetikzlibrary{shapes}

\definecolor{minered}{HTML}{FF0000}
\definecolor{mineblue}{HTML}{4169E1}
\definecolor{minedarkblue}{HTML}{AFEEEE}
\definecolor{minegreen}{HTML}{228B22}
\definecolor{colorCa01}{HTML}{FA9070}
\definecolor{colorCa015}{HTML}{F37457}
\definecolor{colorCa02}{HTML}{EC583E}
\definecolor{colorCa03}{HTML}{CF2613}
\definecolor{colorCa04}{HTML}{7A180C}
\definecolor{colorCa045}{HTML}{511008}
\definecolor{colorCa055}{HTML}{000000}

\newcommand{\Ca}{\text{Ca}}
\newcommand{\Rey}{\text{Re}}

\newcommand{\Cacr}{\text{Ca}_{\mbox{\scriptsize cr}}}

\newcommand{\trace}[1]{\mbox{tr}\left(#1\right)}

\renewcommand{\vec}[1]{\boldsymbol{\bm{#1}}} 

\newcommand{\MM}{\mathrm{MM}}
\newcommand{\QS}{\mathrm{QS}}
\newcommand{\EMM}{\mathrm{EMM}}
\newcommand{\MMCa}{\mathrm{\overline{MM}}}

\title{Deformation of ellipsoidal droplets in homogeneous and isotropic turbulence}
\author{
Fabio Guglietta\\
  Department of Physics \& INFN, University of Rome ``Tor Vergata'', \\ Via della Ricerca Scientifica 1, 00133, Rome, Italy.\\
  \texttt{fabio.guglietta@roma2.infn.it} \\
\And
  Diego Taglienti\\ 
  Department of Physics \& INFN, University of Rome ``Tor Vergata'', \\ Via della Ricerca Scientifica 1, 00133, Rome, Italy.\\
\And
Mauro Sbragaglia\\
Department of Physics \& INFN, University of Rome ``Tor Vergata'', \\ Via della Ricerca Scientifica 1, 00133, Rome, Italy.\\}%

\begin{document}
\maketitle
\begin{abstract}
We study the statistics of deformation of neutrally buoyant droplets in homogeneous isotropic turbulence (HIT), wherein the characteristic droplet size $R$ is smaller than the characteristic Kolmogorov scale $\eta$ of the turbulent flow. We systematically focus on the characterization of droplet statistics obtained with various phenomenological ellipsoidal models (EMs) -- assuming that the droplet preserves the ellipsoidal shape at all times -- with the droplet moving as a passive tracer in the turbulent flow. The predictions of the EMs are compared with ground-truth data obtained with three-dimensional fully resolved simulations (FRSs) without any ad-hoc assumption on the droplet shape. Our work helps in elucidating the applicability of the EMs in describing droplet deformation in HIT at changing the capillary number $\Ca=\tau_{\sigma}/\tau_{\eta}$, weighting the relative importance of the droplet characteristic time $\tau_{\sigma}$ with respect to the turbulent flow characteristic time $\tau_{\eta}$.
\end{abstract}

\maketitle

\section{Introduction}
\begin{figure}[t!]
\centering
\includegraphics[width=1.\linewidth]{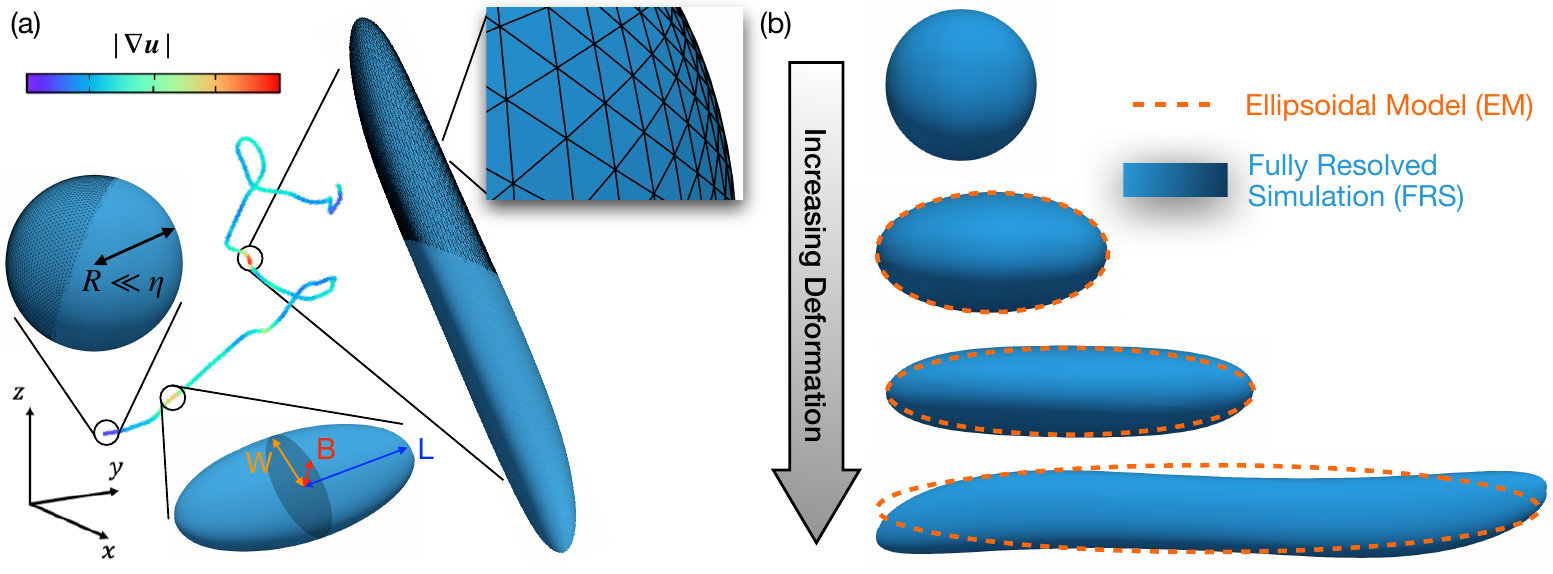}%
\caption{Panel (a): fully resolved simulations (FRSs) of sub-Kolmogorov droplet dynamics along a homogeneous isotropic turbulence (HIT) trajectory in an incompressible flow. The colorbar indicates the local magnitude of the velocity gradient $\boldsymbol{\nabla u}$ along the trajectory. 
FRSs are performed with the numerical methodology described in Ref.~\cite{Taglienti2024}, capable of reproducing droplets with a structured triangular mesh (see zoom on the interface in the center-top panel) at the interface undergoing their dynamics in HIT, wherein the characteristic droplet size ($R$) falls below the characteristic Kolmogorov scale ($\eta$) of the HIT.
Droplet morphology is quantified via the construction of an equivalent ellipsoid with semiaxes $L$, $B$, $W$ ($L>W>B$). Statistics of $L$, $B$ , $W$ over an ensemble of HIT trajectories is computed and compared with corresponding ellipsoidal models (EMs) predictions [see Eq.~\eqref{eq:MMmodel} and text for more details], assuming ellipsoidal shapes at all times.
Panel (b): illustration of the effect of increasing deformation on droplet morphology.}\label{fig:sketch}
\end{figure}

Turbulent flows laden with droplets or bubbles are widely encountered in several settings, including atmospheric clouds~\cite{Shaw2003}, combustion chambers~\cite{carl2001experimental}, nuclear reactors~\cite{poullikkas2003effects}, oceanic transport~\cite{Nissanka2018} and ocean spray aerosols~\cite{villermaux2022bubbles,deike2022mass} just to cite a few examples. In these situations, droplets and bubbles move in a complex fluid dynamics environment and deform due to the action of the outer surrounding flow. Various reviews have been written on the topic~\cite{harper1972motion,rallison1984deformation,stone1994dynamics,fischer2007emulsion,magnaudet2000motion,cristini2004theory,lohse2018bubble,elghobashi2019direct,ni2024deformation}. The process of deformation depends on how the hydrodynamic and surface forces balance at the droplet interface, which we consider in this paper solely characterized by a homogeneous surface tension $\sigma$. Such balance is quantified by the dimensionless capillary number $\Ca=\tau_{\sigma}/\tau_{\eta}$, where $\tau_{\eta}$ is a characteristic time of the flow and $\tau_{\sigma}=\mu R/\sigma$ the characteristic droplet time, with $R$ a representative length scale for the droplet [we choose $R$ as the radius at rest, see Fig.~\ref{fig:sketch}(a)] and $\mu$ the outer flow viscosity. The time $\tau_{\sigma}$ represents the time it takes for the droplet to relax after deformation or, equivalently, to adapt to an outer perturbation with characteristic time $\tau_{\eta}$. If $\tau_{\sigma} \ll \tau_{\eta}$ (i.e., small $\Ca$), we have a linear regime where small deviations from sphericity are promoted by the outer flow; when $\tau_{\sigma} \sim \tau_{\eta}$ (i.e., large $\Ca$), we have a non-linear response resulting in larger deformations [see Fig.~\ref{fig:sketch}(b)]. Furthermore, in the paradigmatic set-up of homogeneous and isotropic turbulence (HIT), the flow is typically generated by energy injection at large scales. Energy is then transported via inertial (non-dissipative) contributions towards the small scales, where the energy spectrum is cut by viscous effects at the Kolmogorov scale $\eta$~\cite{frisch1995turbulence}. Therefore, droplet dynamics crucially depends on the ratio between the droplet scale $R$ and the Kolmogorov length scale $\eta$: for $R \gg \eta$, inertial effects play a role in droplet dynamics whereas, for $R \ll \eta$, they are negligible and one can describe the coupling between the droplet or bubble and the fluid via the Stokes equations~\cite{cristini2003turbulence}. In this paper, we focus on the latter case. 

Due to the complexity of the problem under study, numerical simulations are frequently used to unravel many questions~\cite{elghobashi2019direct}. Different numerical techniques have been employed to investigate the dynamics of sub-Kolmogorov droplets and bubbles, like the boundary integral method~\cite{cristini2003drop,cristini2003turbulence}, the diffuse-interface lattice Boltzmann  method~\cite{milan2020sub}, approaches based on point-like (i.e., non-deformable) particle assumption~\cite{balachandar2010turbulent,kuerten2015effect,russo2014water,ferrante2004physical,mazzitelli2003relevance,le2005direct,mashayek1998droplet,miller1999direct}, 
point-like particle approaches coupled to subgrid models~\cite{spandan2017deformable}, dynamics of neutrally buoyant sub-Kolmogorov droplets~\cite{biferale2014deformation,spandan2016deformation,Ray2018}, hybrid immersed boundary-lattice Boltzmann methods~\cite{Taglienti2024}. In some of these studies~\cite{biferale2014deformation,spandan2016deformation,spandan2017deformable,Ray2018}, instead of directly considering the Stokes equations for the deformable droplets, it is assumed a reduced order model relying on simpler ordinary differential equations regulating the evolution of a morphological tensor $\boldsymbol{S}$ used to describe the droplet interface~\cite{minale2010models}. These models are based on the assumption of small deformations ($\Ca \ll 1$), implying that the droplet retains an ellipsoidal shape at all times~\cite{maffettone1998equation,minale2010models} [see Fig.~\ref{fig:sketch}(b)], hence the name of {\it ellipsoidal models} (EMs). As shown in Refs.~\cite{biferale2014deformation,Ray2018,spandan2016deformation,spandan2017deformable}, EMs can be used to infer relevant features regarding droplet rheology, such as deformation, orientation, morphology, and breakup. Furthermore, this approach allows for efficient statistical analyses, for instance, when considering a collection of Lagrangian trajectories followed and probed with sub-Kolmogorov droplets. In light of experimental observations for stationary flows~\cite{GuidoVillone1997experim,vananroye2006effect,feigl2007}, droplets are found to retain an ellipsoidal shape even for finite to high $\Ca$ values, so that one may ask whether extending the use of EMs in the high capillary regime to complex turbulent flows could be effective to reproduce rheological data. The aim of this paper is to address this question by considering the statistics of neutrally buoyant sub-Kolmogorov droplet deformation in HIT, and comparing results from EMs with those of fully resolved simulations (FRSs). 
{Specifically, we extract the information on the velocity gradients probed from direct numerical simulations of HIT~\cite{biferale2023turb}, and we use it to both numerically integrate the EMs and to perform ab-initio FRSs~\cite{Taglienti2024}, thereby obtaining quantitative information on the droplet’s principal axes of deformation [see Fig.~\ref{fig:sketch}] and enabling a direct comparison between EMs and FRSs.}

The paper is organized as follows: in Sec.~\ref{sec:EM}, FRSs and EMs are reviewed and described; then in Sec.~\ref{sec:methods} we present the numerical framework employed to perform the FRSs; results are presented in Sec.~\ref{sec:results}; finally, conclusions are provided in Sec.~\ref{sec:conclusions}. 


\section{Fully Resolved Simulations and Ellipsoidal Models}\label{sec:EM}
The dynamics of droplets in turbulence can be investigated using two complementary approaches. On the one hand, fully resolved simulations offer high-fidelity, ab-initio solutions by explicitly resolving the fluid–structure interactions at the droplet interface.
On the other hand, reduced-order models provide an efficient way to approximate droplet deformation based on simplified physics. In the following, we briefly describe both approaches used in this study.

\subsection*{Fully Resolved Simulation (FRSs)}
FRSs performed with the immersed boundary-lattice Boltzmann method~\cite{book:kruger} provide a powerful tool to generate ground-truth data for droplet dynamics in HIT. Such methodology has been recently employed to simulate the dynamics of droplets and soft particles in a variety of setups~\cite{li2019finite,taglienti2023reduced,pelusi2023sharp,guglietta2024analytical,Taglienti2024,bellantoni2025immersed,guglietta2020effects,guglietta2021loading,guglietta2020lattice,Guglietta2023}. In particular, for the simulations of droplets in turbulence, we employ the numerical method described in Ref.~\cite{Taglienti2024}. In a nutshell, the lattice Boltzmann method is used to simulate the fluid both inside and outside the droplet, and the immersed boundary method couples the fluid with the droplet interface, the latter being represented by a triangular mesh [see Fig.~\ref{fig:sketch}(a)] and characterized by a tunable surface tension $\sigma$.
Numerical simulations are performed with a mesh-smoothing technique to regularize the triangular mesh~\cite{Taglienti2024}. We refer the reader interested in technical details to Ref.~\cite{Taglienti2024}.

\subsection*{Ellipsoidal models (EMs)}
EMs are reduced-order models designed to describe the evolution of droplet shape, which is represented by a second-order morphological tensor $\boldsymbol{S}$. These models offer two main advantages. First, the evolution equation for $\boldsymbol{S}$ is governed solely by the velocity gradient tensor $\nabla\boldsymbol{u}$ and its interactions with $\boldsymbol{S}$, without requiring full knowledge of the flow field. Second, the resulting ordinary differential equation can be integrated efficiently using standard numerical methods, such as Runge–Kutta schemes, instead of performing ab-initio numerical simulations. It is important to stress that EMs assume the droplet to be always ellipsoidal: while this is a good approximation for small degrees of deformation, it does not hold anymore when the deformation is large, e.g., close to the breakup [see Fig.~\ref{fig:sketch}(b)].  Information about droplet deformation can be obtained from the eigenvalues of the tensor $\boldsymbol{S}$, which correspond to the squares of the ellipsoid’s semiaxes. Taking the square roots of these eigenvalues yields the lengths of the droplet’s principal semiaxes, which we denote as the length ($L$), breadth ($B$), and width ($W$), conventionally ordered such that $L > W > B$ (see Fig.~\ref{fig:sketch}).
\subsection{Maffettone and Minale ($\MM$) model.}\label{sec:EM_MM}
One of the most celebrated EMs is that proposed by Maffettone and Minale (MM)~\cite{maffettone1998equation}. The model equation describing the time evolution of $\boldsymbol{S}$ is:
\begin{equation}\label{eq:MMmodel}
\frac{d\boldsymbol{S}}{dt}-\left(\boldsymbol{\Omega}\cdot\boldsymbol{S}-\boldsymbol{S}\cdot\boldsymbol{\Omega}\right)= -\frac{f_1}{\tau_{\sigma}} \left[\boldsymbol{S}-g\left(\boldsymbol{S}\right)\boldsymbol{I}\right]+f_2 \left(\boldsymbol{E}\cdot\boldsymbol{S}+\boldsymbol{S}\cdot\boldsymbol{E}\right)\ ,
\end{equation}
where 
\begin{equation}
\boldsymbol{\Omega}=\frac{1}{2}\left(\boldsymbol\nabla\vec{u}-\boldsymbol\nabla\vec{u}^T\right) ,\quad\quad
\boldsymbol{E}= \frac{1}{2}\left(\boldsymbol\nabla\vec{u}+\boldsymbol\nabla\vec{u}^T\right)
\end{equation}
are the vorticity and the rate-of-strain tensors, respectively. The non-linear function $g\left(\boldsymbol{S}\right)=\frac{3 III_S}{II_S}$ 
is used to enforce volume conservation, with $II_S=\frac{1}{2}\left[\trace{\boldsymbol{S}}^2-\trace{\boldsymbol{S}^2}\right]$ and $III_S=\det(\boldsymbol{S})$ being the second and third invariant of the tensor $\boldsymbol{S}$, respectively~\cite{maffettone1998equation}. The two parameters $f_1=f_1(\lambda)$ and $f_2=f_2(\lambda)$ are set by the viscosity ratio between the inner and outer fluid, $\lambda$, and are retrieved in the framework of the perturbation theory for small deformations ($\Ca \ll 1$) applied to Stokes equations~\cite{rallison1980note,maffettone1998equation}.
Indeed, expanding the morphological tensor $\boldsymbol{S}$ up to second order (excluded) in capillary number: $\boldsymbol{S}=\boldsymbol{I} R^2+ \boldsymbol{S^{(1)}}\Ca+{\cal O}(\Ca^2)$, and plugging this result in Eq.~\eqref{eq:MMmodel}, we obtain~\cite{maffettone1998equation}:
\begin{equation}\label{eq:MM_linear}
    \frac{d\boldsymbol{S}}{dt} = -\frac{f_1}{\tau_\sigma}\left[\boldsymbol{S}-\boldsymbol{I}R^2\right]+2 R^2f_2\boldsymbol{E}\ ,
\end{equation}
where terms of ${\cal O}(\Ca^2)$ have been neglected. We will refer to the model descibed by Eq.~\eqref{eq:MM_linear} as linear theory (LT). By comparing Eq.~\eqref{eq:MM_linear} with the analytical results in the perturbation theory aforementioned~\cite{rallison1980note,maffettone1998equation}, Maffettone and Minale obtained~\cite{maffettone1998equation}:
\begin{equation}\label{eq:f1f2mm}
f^{\tiny\MM}_1(\lambda)=\frac{40\left(\lambda+1\right)}{\left(2\lambda+3\right)\left(19\lambda+16\right)}\ ,\quad\quad f^{\tiny\MM}_2(\lambda)=\frac{5}{2\lambda+3}\ .
\end{equation}
We remark that the model in Eq.~\eqref{eq:MMmodel} exploits the linear nature of the Stokes equations to superimpose the deformation effects given by the flow with the counteracting surface tension effects; 
the function $g(\boldsymbol{S})$ cannot be derived from linear theory, and is thus manually inserted into the model to enforce volume conservation by requiring the determinant of $\boldsymbol{S}$ to be constant~\cite{maffettone1998equation}. 
{It is worth noting that the LT model has a trivial volume conservation that reads $g(\boldsymbol{S}) = R^2$.}
This conservation is essential according to the primitive equations, where the divergence of the velocity field is zero.
\subsection{Modified Maffettone and Minale ($\MMCa$)}\label{sec:EM_MMCa}
Maffettone and Minale noted that the parameter $f_2=f^{\tiny\MM}_2$ defined in Eq.~\eqref{eq:f1f2mm} underestimates the deformation and overestimates the values of the critical capillary number $\Cacr$ (i.e., the capillary number at which the droplet shows breakup), which can be computed as a function of $f_1$ and $f_2$~\cite{maffettone1998equation}. For such reasons, they propose a different choice of $f_2$, allowing the latter to depend also on $\Ca$ (we will refer to this model extension as $\MMCa$ model): 
\begin{equation}\label{eq:f1f2mmca}
    f^{\tiny{\MMCa}}_2(\lambda,\Ca)=\frac{5}{2\lambda+3} + \frac{3\Ca^2}{2+6\Ca^{2+\delta}}\frac{1}{1+\epsilon\lambda^2}\ ,
\end{equation}
with $\delta$ and $\epsilon$ being small positive numbers. However, Maffettone and Minale stated that if $\lambda$ and $\Ca$ are not large, then $\delta=\epsilon=0$ (we adopt the same choice in this work). Note that $f_2^{\tiny{\MMCa}}\to f^{\tiny{\MM}}_2$ in the limit $\Ca\to 0$. 
\subsection{Extended Maffettone and Minale ($\EMM$).}
We already showed that Maffettone and Minale~\cite{maffettone1998equation} matched the parameters $f_1=f^{\tiny\MM}_1$ and $f_2=f^{\tiny\MM}_2$ with the analytical results obtained in the perturbation theory~\cite{maffettone1998equation,rallison1984deformation,rallison1980note}, thus making them intrinsically built on the assumption of small deformation. 
Indeed, the MM model underestimates the deformation, in particular for high values of capillary number $\Ca$, even for steady flows~\cite{maffettone1998equation,taglienti2023reduced}. 
In our previous work~\cite{taglienti2023reduced}, we proposed an extension of the MM model ($\EMM$) by providing a polynomial expression in $\Ca$ for $f_1$ and $f_2$:
\begin{equation}\label{eq:f1f2emm}
    f_{1,2}^{\tiny\EMM}(\lambda,\Ca) = \sum_ia_i^{(1,2)}(\lambda) \Ca^i\ ,
\end{equation}
where the coefficients $a_i^{(1,2)}$ were fitted with data from numerical simulations. Values of $a_i^{(1,2)}$ for $\lambda=1$ are reported here~\footnote{The coefficients $a_i^{(1,2)}(\lambda=1)$ for $i\in[0,5]$ are reported:
$a_0^{(1)}=0.457$, 
$a_1^{(1)}=0.235$, 
$a_2^{(1)}=-4.546$, 
$a_3^{(1)}=20.536$, 
$a_4^{(1)}=-34.798$; and  
$a_0^{(2)}=1$, and $a_i^{(2)}=0$ for $i\in[1,5]$.}.
\subsection{Quasi-static ($\QS$) Approximation.}
\begin{figure}[t!]
\centering
\includegraphics[width=0.6\linewidth]{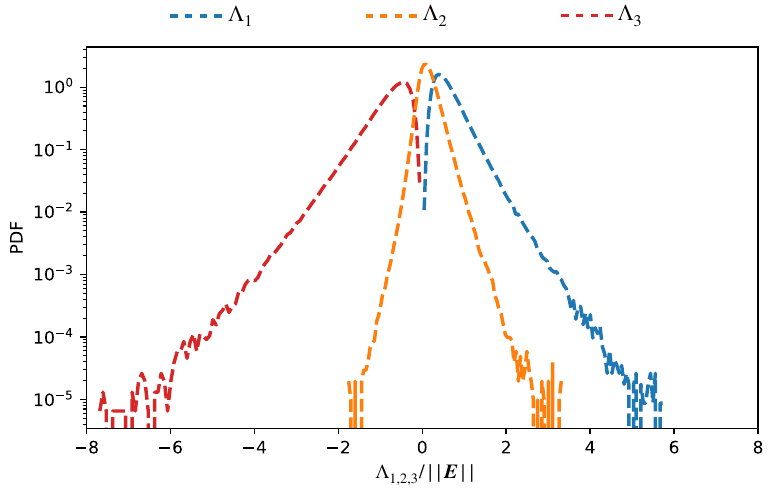}%
\caption{PDFs of the eigenvalues $\Lambda_{1,2,3}$ of the rate of strain tensor $\vec{E}$ of the HIT turbulent trajectories used for the analysis of droplet deformation. Eigenvalues are normalized with the norm $||\vec{E}||$ of the rate-of-strain tensor $\vec{E}$.}\label{fig:Eigen_E}
\end{figure}

In the limit of small deformations (i.e., $\Ca \rightarrow 0$), the droplet dynamics is expected to be quasi-static (QS), $d \boldsymbol{S}/ dt \approx 0$, due to the fact that the characteristic droplet time is much smaller than the characteristic time of the outer flow~\cite{cristini2003drop}. From Eq.~\eqref{eq:MMmodel}, in the limit $\Ca \rightarrow 0$, we have 
$g\left(\boldsymbol{S}\right) \rightarrow R^2$, 
$\left(\boldsymbol{\Omega}\cdot\boldsymbol{S}-\boldsymbol{S}\cdot\boldsymbol{\Omega}\right) \rightarrow 0$ and 
$\left(\boldsymbol{E}\cdot\boldsymbol{S}+\boldsymbol{S}\cdot\boldsymbol{E}\right) \rightarrow 2 R^2\boldsymbol{E}$, so that QS approximation implies:
\begin{equation}\label{eq:QS}
\boldsymbol{S}_{\text{QS}} = \boldsymbol{I}R^2+\frac{2 f_2 \tau_{\sigma}}{f_1} R^2\boldsymbol{E} \ .
\end{equation}
In other words, in the limit $\Ca \rightarrow 0$, there is no need to integrate Eq.~\eqref{eq:MMmodel}, since the eigenvalues of the morphological tensor $\boldsymbol{S}$ can be reconstructed from the eigenvalues $\Lambda_{1,2,3}$ of the rate-of-strain tensor $\boldsymbol{E}$ (see Fig.~\ref{fig:Eigen_E}).

\begin{figure}[t!]
\centering
\includegraphics[width=1.\linewidth]{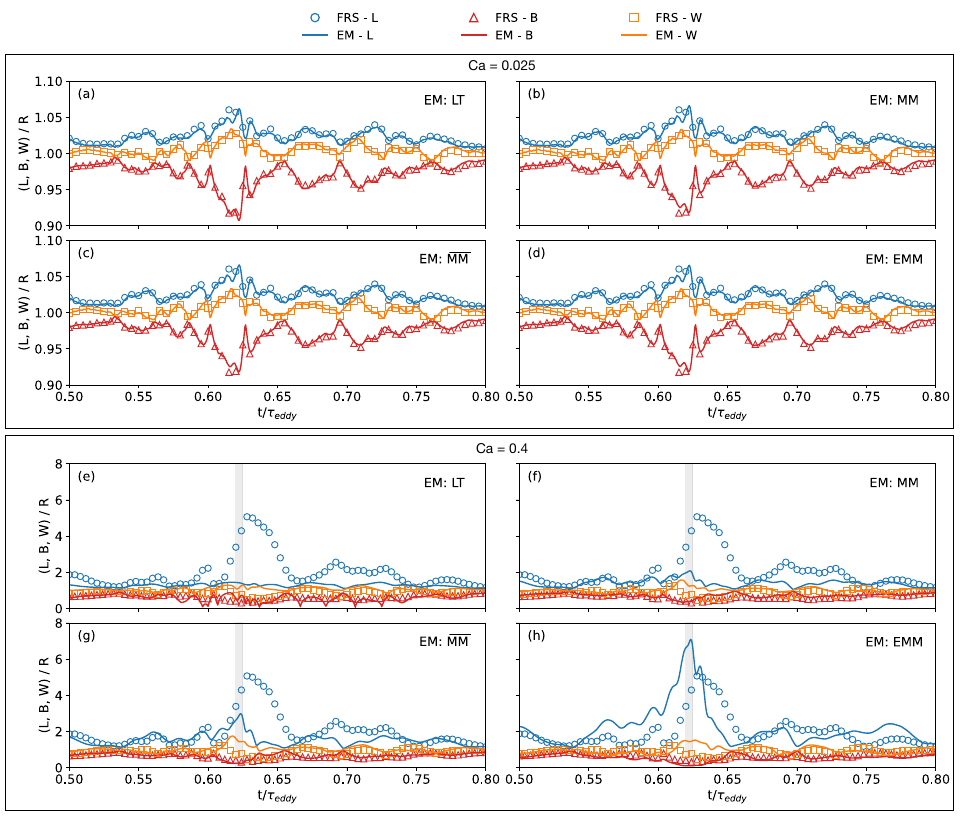}%
\caption{Comparison between {fully resolved simulations (FRSs) and different ellipsoidal models (EMs)} for two values of capillary number, $\Ca = 0.025$ (top panels) and $\Ca = 0.4$ (bottom panels), along a representative turbulent trajectory. Panels (a–d) correspond to $\Ca = 0.025$, showing the normalized semiaxes $L/R$, $B/R$, and $W/R$ as functions of normalized time $t/\tau_{\text{eddy}}$, as predicted by: (a) the linear theory (LT), (b) the Maffettone–Minale (MM) model, (c) the modified MM model with $f_2$ given in Eq.~\eqref{eq:f1f2mmca} ($\MMCa$), and (d) the extended MM model (EMM). Panels (e–h) show the same analysis for $\Ca = 0.4$, where droplet deformation is significantly more pronounced. In each panel, symbols represent FRSs data (blue circles for $L$, red triangles for $B$, and yellow squares for $W$), while solid lines show the predictions from the corresponding EM. The vertical gray band highlights a region of peak deformation (see Sec.~\ref{sec:results}).} \label{fig_8panels}
\end{figure}

\begin{figure}[t!]
\centering   
\includegraphics[width=1.\linewidth]{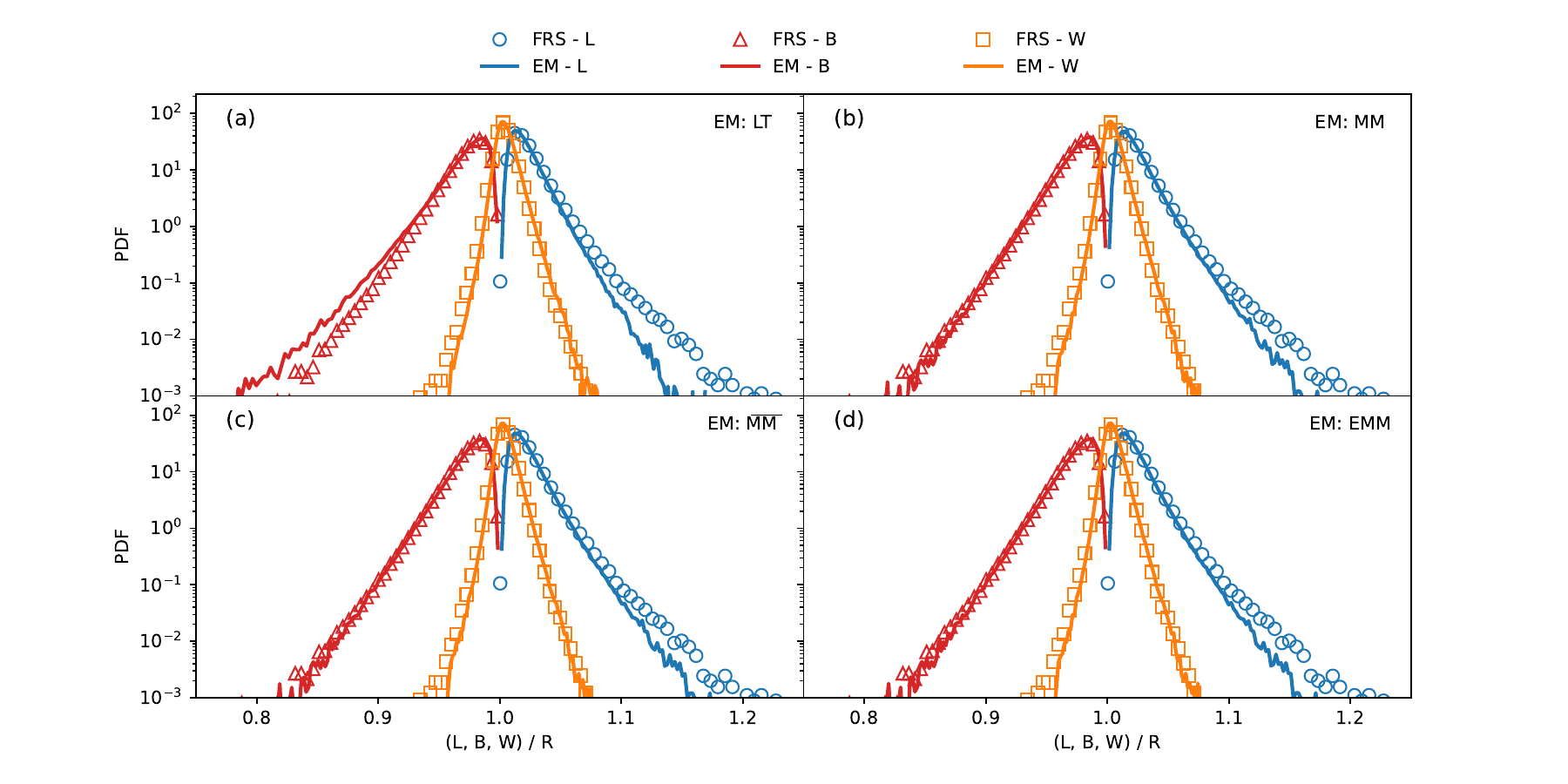}
\caption{Probability distribution functions (PDFs) of normalized droplet semiaxes $L/R$, $B/R$, and $W/R$ for a statistical ensemble of turbulent trajectories at $\Ca = 0.025$. The statistics is collected from an ensemble of 1000 independent droplet trajectories in HIT.
Panels (a–d) compare the {fully resolved simulation (FRS)} data (symbols) with the predictions of various ellipsoidal models {(EMs)} (solid lines):  (a) the linear theory (LT), (b) the Maffettone–Minale (MM) model, (c) the modified MM model with $f_2$ given in Eq.~\eqref{eq:f1f2mmca} ($\MMCa$), and (d) the extended MM model (EMM). 
Colors and markers denote the three principal semiaxes: blue circles and lines for $L$, red triangles and lines for $B$, and yellow squares and lines for $W$. All PDFs are shown on a logarithmic scale to emphasize tail behavior.}\label{fig:result_lowCa}
\end{figure}
\section{Numerical Setup}\label{sec:methods}
To assess the accuracy of the EMs introduced in Sec.~\ref{sec:EM} within the context of HIT, we compare their predictions against FRSs under identical flow conditions. In both cases, droplet dynamics are driven by the same set of time-dependent velocity gradients $\boldsymbol{\nabla u}(t)$, extracted from direct numerical simulations of HIT~\cite{biferale2023turb}. The focus is on quantifying the statistical features of droplet deformation across a range of capillary numbers and evaluating the impact of different closure strategies for the EM model coefficients $f_1$ and $f_2$. The HIT flow field is characterized by a Taylor-scale Reynolds number $\Rey \sim 130$ and is drawn from the TURB-Lagr open-source database~\cite{biferale2023turb}, which provides over $3 \times 10^5$ particle trajectories. 
Each trajectory lasts for roughly an eddy turnover time $\tau_{\text{eddy}} \sim 200\tau_\eta$~\cite{biferale2023turb}.
For completeness, in Fig.~\ref{fig:Eigen_E}, we provide details on the PDF of the eigenvalues $\Lambda_{1,2,3}$ of the rate-of-strain tensor $\boldsymbol{E}$ extracted from the HIT trajectories that we utilized~\cite{kerr1987histograms,tsinober1992experimental,lund1994improved,su1996scalar,tsinober2009informal,tom2021exploring,buaria2022generation,carbone2024asymptotic}. Consistently with previous literature results, we have positively/negatively skewed PDFs for the maximum/minimum eigenvalues $\Lambda_{1,3}$, and a positively skewed PDF for the intermediate eigenvalue $\Lambda_2$, this latter fact being crucially related to a non-zero strain production of the turbulent flow~\cite{luthi2005lagrangian}.
Following earlier investigations, we treat neutrally buoyant droplets with a unitary viscosity ratio ($\lambda=1$) and a unitary density ratio between the droplet fluid and the outer fluid. Droplets are treated as passive tracers, following the Lagrangian trajectories of an outer turbulent flow. The outer flow of the FRS domain is adapted to accommodate the turbulent gradient matrix $\partial_i u_j (t)$ at all times under the assumption that we are dealing with a sub-Kolmogorov droplet, hence the outer flow matches a linear flow constructed from the turbulent gradients~\cite{cristini2003turbulence,biferale2014deformation}. The analysis for droplet deformation is conducted for three different capillary numbers: $\Ca=0.025$, $\Ca=0.15$, and $\Ca=0.4$. For each $\Ca$ studied,  simulations for one thousand independent trajectories are performed. Concerning the FRSs, the fluid domain consists of a cubic box with size $L_x=L_y=L_z=10 R$; the droplet is simulated with a structured mesh composed of 20480 triangles (see Fig.~\ref{fig:sketch}). The radius of the droplet at rest is chosen as $R/\eta=0.472$, which guarantees a fair convergence to the sub-Kolmogorov dynamical equations~\cite{Taglienti2024}. 
Since FRSs are performed within a finite fluid domain, a threshold is imposed on the droplet elongation: when $L/R$ exceeds 5, the corresponding trajectory is excluded from the statistics and replaced by a new one (this effect is particularly prominent at high capillary numbers, such as $\Ca = 0.4$). Such a condition mimics the onset of breakup in finite domains, consistent with observations from Ref.~\cite{cristini2003turbulence}. 
All simulations are run using an in-house GPU-accelerated code on Nvidia A100 GPUs, each requiring roughly three hours of computation time. To generate EM data, we use the same set of turbulent trajectories from the HIT database. Instead of resolving the fluid–structure interaction, we integrate the tensorial evolution Eq.~\eqref{eq:MMmodel}, where the time-dependent tensors $\boldsymbol{E}(t)$ and $\boldsymbol{\Omega}(t)$ are dynamically updated along each trajectory. Integration is performed with a standard explicit Runge–Kutta scheme~\cite{Butcher1996}. We consider four EMs: the linear theory (LT) model [Eq.~\eqref{eq:MM_linear}], the original MM model [Eqs.~\eqref{eq:MMmodel},\eqref{eq:f1f2mm}], the $\MMCa$ model with modified $f_2$ [Eq.~\eqref{eq:f1f2mmca}], and the extended model $\EMM$ [Eq.~\eqref{eq:f1f2emm}]. These comparisons aim to evaluate how each model captures the deformation statistics, particularly across different regimes of $\Ca$, and clarify the role of nonlinearities and volume conservation in improving model fidelity.

\begin{figure}[t!]
\centering
    \includegraphics[width=0.7\linewidth]{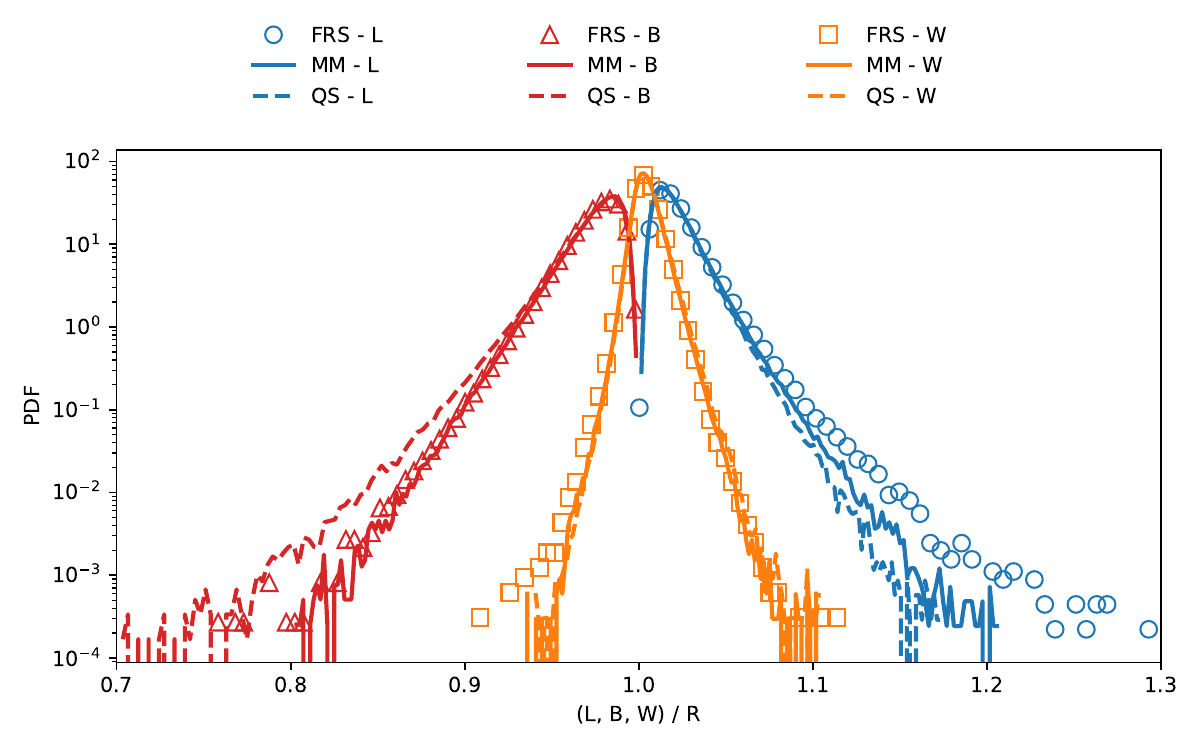}
\caption{Probability distribution functions (PDFs) of normalized droplet semiaxes $L/R$, $B/R$, and $W/R$ for a statistical ensemble of turbulent trajectories at $\Ca = 0.025$. The statistics is collected from an ensemble of 1000 independent droplet trajectories in HIT. Comparison between the {fully resolved simulation (FRS)} data (symbols) with the predictions of the Maffettone and Minale (MM) model (solid lines) and the quasi-static (QS, dashed lines) approximation (see Eq.~\ref{eq:QS}).
Colors and markers denote the three principal semiaxes: blue circles and lines for $L$, red triangles and lines for $B$, and yellow squares and lines for $W$. All PDFs are shown on a logarithmic scale to emphasize tail behavior.}\label{fig:QSprediction}
\end{figure}

\begin{figure}[t!]
\centering
\includegraphics[width=1.0\linewidth]{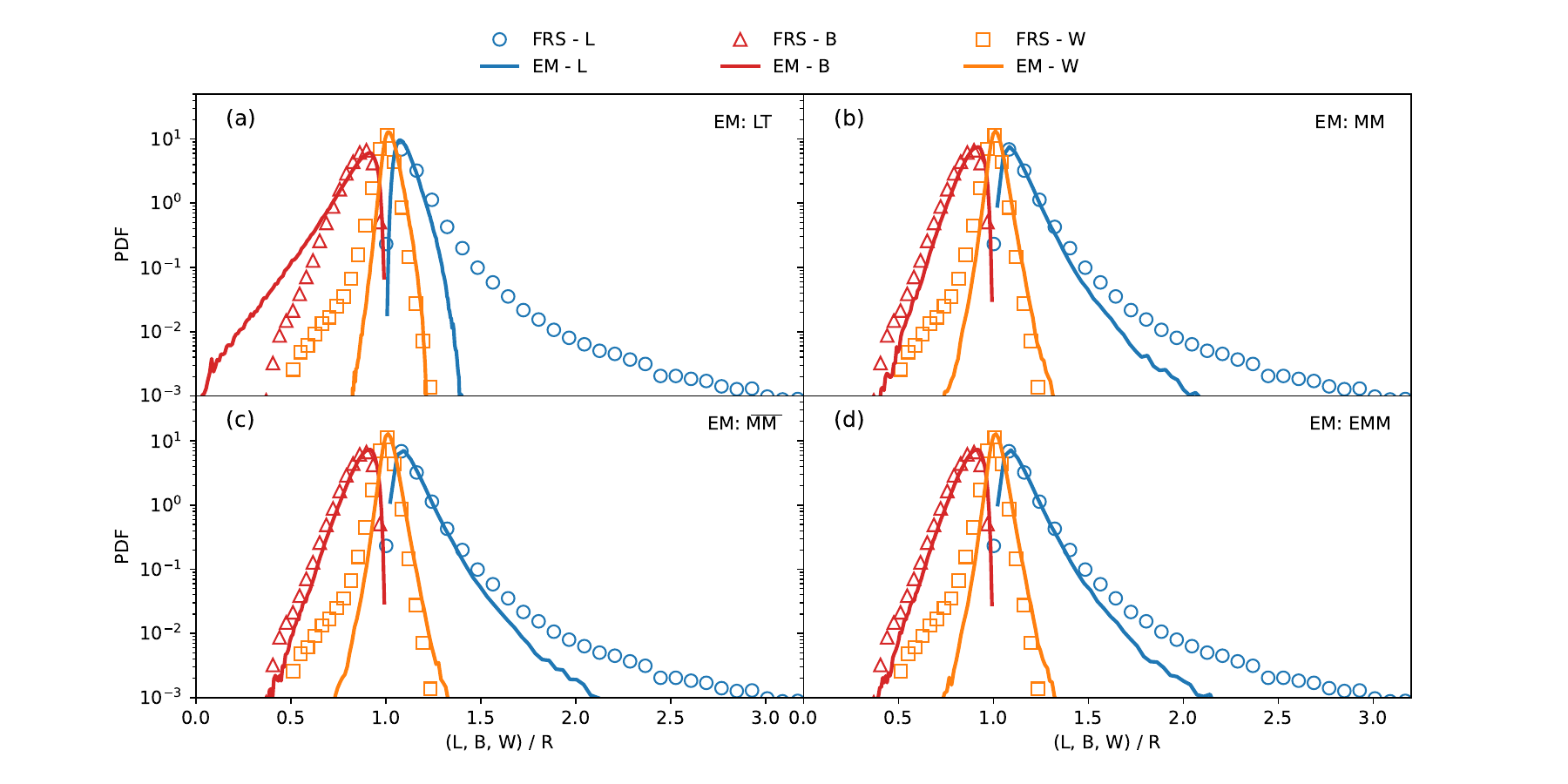}%
\caption{Probability distribution functions (PDFs) of normalized droplet semiaxes $L/R$, $B/R$, and $W/R$ for a statistical ensemble of turbulent trajectories at $\Ca = 0.15$. The statistics is collected from an ensemble of 1000 independent droplet trajectories in HIT.
Panels (a–d) compare the {fully resolved simulation (FRS)} data (symbols) with the predictions of various ellipsoidal models {(EMs)} (solid lines): (a) the linear theory (LT), (b) the Maffettone–Minale (MM) model, (c) the modified MM model with $f_2$ given in Eq.~\eqref{eq:f1f2mmca} ($\MMCa$), and (d) the extended MM model (EMM). 
Colors and markers denote the three principal semiaxes: blue circles and lines for $L$, red triangles and lines for $B$, and yellow squares and lines for $W$. All PDFs are shown on a logarithmic scale to emphasize tail behavior.}\label{fig:result_MediumCa}
\end{figure}

\begin{figure}[t!]
\centering
\includegraphics[width=1.0\linewidth]{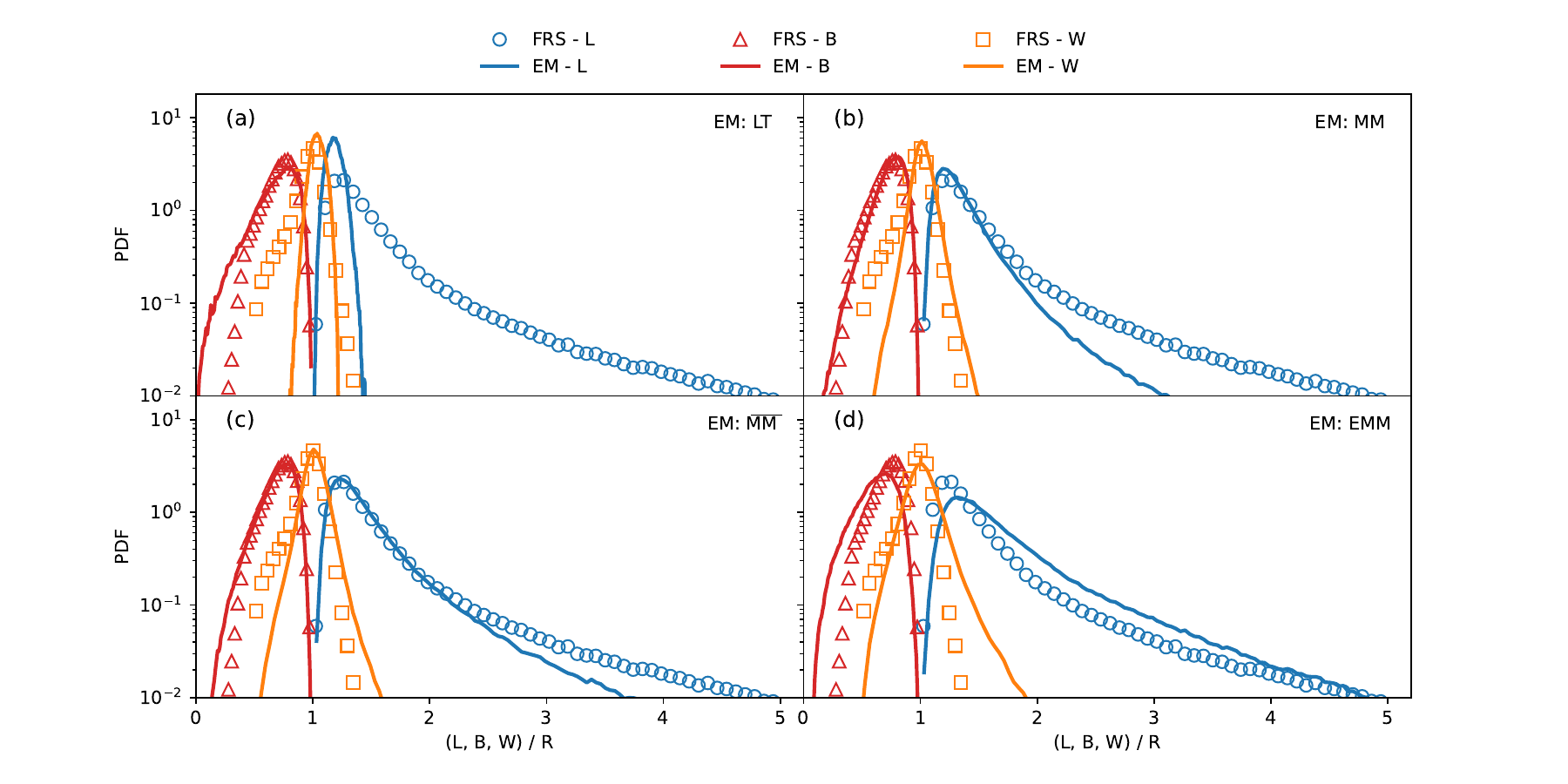}%
\caption{Same as Fig.~\ref{fig:result_MediumCa}, for $\Ca=0.4$.}\label{fig:result_HighCa}
\end{figure}

\section{Results}\label{sec:results}
In Fig.~\ref{fig_8panels}, we analyze a representative turbulent trajectory and report the {time evolution of the} normalized droplet semiaxes, $L/R$, $B/R$, and $W/R$ [see also Fig.~\ref{fig:sketch}(a)], for the smallest and largest investigated capillary numbers $\Ca$. {Time is normalized with the eddy turnover time $\tau_{\text{eddy}}$~\cite{biferale2023turb}.}
Data from FRSs are reported with markers, while data from EMs are reported with solid lines. Different colors represent different semiaxes. At $\Ca = 0.025$, all EMs exhibit excellent agreement with the FRSs. In this regime, deviations from sphericity are minimal (i.e., $\boldsymbol{S} \rightarrow \boldsymbol{I}R^2$), and even the LT prediction [see Eq.~\eqref{eq:MM_linear}] provides accurate results.
The small $\Ca$ limit implies that the droplet relaxation time is much shorter than the characteristic timescale of the flow, making the coupling between $\boldsymbol{\Omega}$ and $\boldsymbol{S}$, which appears on the l.h.s. of Eq.~\eqref{eq:MMmodel} and is known to scale as ${\cal O}(\Ca^2)$~\cite{rallison1980note,maffettone1998equation}, negligible.  Notably, the LT model can be derived analytically from the Stokes equations in the asymptotic limit $\Ca \rightarrow 0$~\cite{rallison1980note}.
Thus, the analysis at small $\Ca$ also serves as a benchmark for the numerical framework, validating both the accuracy of the FRSs and extending the validation of Ref.~\cite{Taglienti2024} to include all droplet semiaxes. 
For $\Ca=0.4$, data from FRSs reveal (see $t/\tau_{\text{eddy}} \approx 0.62$, in correspondence of the gray-shaded region) a correlation between enhanced peaks in $L/R$ and sudden compressions on the orthogonal semiaxes ($W$ and $B$). Such behavior was also observed in Ref.~\cite{biferale2014deformation}, where the authors identified a transition first from an oblate shape (i.e., $L/R\sim W/R$ and $B/R<1$) to a prolate shape (i.e., $L/R\gg 1$ and $W/R\sim B/R<1$), and then the droplet recovers the spherical shape (i.e., $L/R\sim W/R\sim B/R\sim 1$). It is worth noting that, in the specific example of Fig.~\ref{fig_8panels}, among the EMs considered, the $\MMCa$ and $\EMM$ models best capture the high values of $L/R$. In particular, $\MMCa$ tends to slightly underestimate the FRS data, while $\EMM$ slightly overestimates it. This trend will also be observed in the statistical analysis reported below (see later Fig.~\ref{fig:result_HighCa}).
We also stress that when the droplets undergo strong deformations, the ellipsoidal assumption does not hold anymore, as can be seen in Fig.~\ref{fig:sketch}, panel (b): the reported droplet shapes come from the FRSs, while the orange dashed lines serve as a reference of the ellipsoidal shape. For $\Ca=0.4$, variations in droplet length $L$ are not strongly pronounced in the LT model nor in the MM and $\MMCa$ models, if seen in comparison to FRS data; among all EMs analyzed, data from $\EMM$ model display more pronounced variations in $L$, a fact that we can attribute on the way the coefficients $f^{\tiny\EMM}_1$ and $f^{\tiny\EMM}_2$ are computed (see Sec.~\ref{sec:EM}). 
Overall, apart from rare events during which the droplet undergoes quick, strong deformations, there are time lapses over the eddy turnover time analyzed where the MM, $\MMCa$, and $\EMM$ predictions show a good agreement with FRSs. We remark, however, that data reported in Fig.~\ref{fig_8panels} just refer to a single turbulent trajectory; hence, it is dutiful to assess the robustness of these findings via some statistical analysis on the various observables on an ensemble of droplet trajectories.

We then study the probability distribution functions (PDFs) of the ellipsoid normalized semiaxes $L/R$, $B/R$, and $W/R$, collecting data for 1000 independent trajectories at changing $\Ca$.  In Fig.~\ref{fig:result_lowCa}, we report data for $\Ca=0.025$. Apart from rare events in the tails, we observe a good matching between the PDF extracted from the FRSs and those extracted from the EMs; thus, the agreement observed in Fig.~\ref{fig_8panels} for $\Ca=0.025$ and for a single trajectory is rather robust throughout the ensemble of turbulent trajectories. The tails of the distributions for $L/R$ and $B/R$ predicted by the LT model exhibit larger discrepancies compared to those from the other EMs. As discussed in Sec.~\ref{sec:EM_MM}, the LT model can be derived from the MM model [see Eq.~\eqref{eq:MMmodel}] when the capillary number approaches zero ($\Ca \to 0$). Moreover, we remark that the LT model has a trivial volume conservation in the context of the EMs, i.e., $g(\boldsymbol{S}) = \boldsymbol{I}R^2$ [see Eq.~\eqref{eq:MM_linear}]. Although data in Fig.~\ref{fig:result_lowCa} corresponds to a relatively small value of $\Ca$ (i.e., $\Ca = 0.025$), the observed deviations suggest that  the effects of a finite capillary number start to emerge. We then expect that the LT model will be the first among the considered EMs to lose predictive accuracy as $\Ca$ increases. As already stated in Sec.~\ref{sec:EM}, in the limit of small deformations ($\Ca \rightarrow 0$), the droplet dynamics is expected to be quasi-static [see Eq.~\eqref{eq:QS}]. This behavior is quantitatively confirmed in Fig.~\ref{fig:QSprediction}, where we report the PDFs of the normalized semiaxes for the FRSs and the MM model (previously shown in Fig.~\ref{fig:result_lowCa}), along with the predictions obtained from the morphological tensor in Eq.~\eqref{eq:QS}, namely $L_{\text{QS}}/R$, $W_{\text{QS}}/R$, and $B_{\text{QS}}/R$. The excellent agreement among these datasets supports the validity of the QS approximation in this low-$\Ca$ regime. {Since in the QS limit, the morphological tensor describing the deviation from sphericity is the strain matrix $\boldsymbol{E}$ multiplied by a factor [i.e., $\boldsymbol{S}_{\text{QS}}-\boldsymbol{I}R^2\sim \boldsymbol{E}$, see Eq.~\eqref{eq:QS}], then the droplet deformation statistics inherits the structural features of the turbulent strain field. In other words, the morphological tensor $\boldsymbol{S}_{\text{QS}}$ becomes ``enslaved'' to the strain matrix $\boldsymbol{E}$. In particular, the PDFs of the normalized semiaxes reflect the positive skewness associated with the intermediate eigenvalue $\Lambda_2$ of the rate-of-strain tensor $\bm{E}$, as highlighted in Fig.~\ref{fig:Eigen_E}.  
The fact that this statistical feature reappears in the droplet deformation analysis underscores a meaningful connection between turbulent strain dynamics and the behavior of sub-Kolmogorov deformable droplets.} It is worth noting that the tails of the QS model for $B/R$ and $L/R$ show similar discrepancies to those of the LT model at $\Ca = 0.025$ [see Fig.~\ref{fig:result_lowCa}, panel (a)]. Indeed, both the QS and LT models are valid in the $\Ca\to 0$ limit.

In Figs.~\ref{fig:result_MediumCa} and~\ref{fig:result_HighCa} we extend our investigation on the PDFs at higher $\Ca$ regime showing results for $\Ca=0.15$ and $\Ca=0.4$, respectively. Regarding the statistics of the normalized length $L/R$, Fig.~\ref{fig:result_MediumCa} (i.e., $\Ca = 0.15$) shows that the LT model [panel (a)] significantly underestimates the probability of large deformations compared to FRS. This limitation becomes even more pronounced at $\Ca = 0.4$ [Fig.~\ref{fig:result_HighCa}, panel (a)], where the MM model [panel (b)] also begins to deviate substantially in the tail, failing to capture extreme elongations. In contrast, both the $\MMCa$ [panel (c)] and EMM [panel (d)] models follow the FRS data more closely, especially in reproducing the long-tailed distribution of $L/R$, demonstrating improved robustness at higher deformation regimes. For the normalized breadth $B/R$, the discrepancies between EMs and FRS data are less severe than those observed for $L/R$: while the LT model still shows visible deviations at both $\Ca = 0.15$ and $\Ca = 0.4$, the other EMs maintain overall a better alignment with the FRS statistics. Concerning the normalized width $W/R$, the evolution of the PDF shape as $\Ca$ increases provides additional insight. As remarked earlier, at lower $\Ca$, the distribution shows a slight positive skewness, consistent with the QS regime and the inherent asymmetry of the turbulent strain field (see Fig.~\ref{fig:Eigen_E}). As $\Ca$ increases, however, the skewness of $W/R$ reverses and becomes negative. This transition indicates the growing influence of large deformation events associated with configurations approaching breakup, where droplets acquire cigar-like topologies elongated in the direction of maximum extension. Such behavior has been reported in previous works~\cite{biferale2014deformation}, where negative skewness in the distribution of minor semiaxis has been linked to precursors of breakup events. At $\Ca = 0.4$, the negative skewness of $W/R$ becomes prominent in the FRS data, yet it is poorly captured by the LT model. The MM model offers a partial improvement, while the $\MMCa$ and EMM models provide a closer approximation of the observed asymmetry. Importantly, at $\Ca = 0.4$, FRS data reveal that droplet elongation can reach values large enough to exceed the domain boundaries. For this reason, in the case of the FRS, a threshold on droplet deformation must be introduced: without it, excessively deformed droplets would exceed the boundaries of the finite-size numerical domain. This issue does not arise for EMs, which are based on the integration of ordinary differential equations and do not involve spatial confinement, allowing, in principle, for unbounded deformation. To ensure a fair comparison, we applied the same deformation threshold to both FRS and EM data. Once this threshold is enforced, we observe that FRS trajectories exceed it more frequently than those predicted by the MM model. However, this is not generally true for all EMs, as the EMM model exhibits threshold-crossing events at a frequency comparable to that of FRSs. This enhanced stability shown by the MM model is consistent with classical findings on droplet breakup thresholds. For example, in simple shear flow with viscosity ratio $\lambda = 1$, breakup is observed at $\Ca_{\text{cr}} \approx 0.41$, whereas the MM model predicts no breakup, i.e., $\Ca_{\text{cr}} = \infty$~\cite{maffettone1998equation}. 
Similarly, in axisymmetric extensional flow, numerical simulations report $\Ca_{\text{cr}} \approx 0.06$~\cite{singh2022numerical}, while the MM model yields a significantly higher threshold, $\Ca_{\text{cr}} \approx 0.228$~\cite{maffettone1998equation}. 
These results confirm that the MM model systematically underestimates the onset of breakup, yielding artificially stable droplet dynamics under turbulent forcing.

\section{Conclusions}\label{sec:conclusions}
We have studied the statistics of sub-Kolmogorov droplet deformation in homogeneous isotropic turbulence (HIT).  With reference to fully resolved simulations (FRSs) performed with the numerical methodology described in Ref.~\cite{Taglienti2024}, the primary focus of our analysis was on the assessment of the quality of the statistics extracted from some reduced order models that are built on the assumption that the droplet is always ellipsoidal -- and for this reason they are referred as ellipsoidal models (EMs).
Among them, we started from the Maffettone and Minale (MM) model, and then we considered other models that can be derived from the MM one, such as: the linear theory (LT) model, which is retrieved in the assumption of $\Ca\to 0$; the model proposed again by Maffettone and Minale~\cite{maffettone1998equation} but with a different choice of $f_2$ in order to make the model able to predict better the critical capillary number (which we referred as $\MMCa$ model); the extension of the Maffettone and Minale (EMM) model proposed by Taglienti~{\it et al.}~\cite{taglienti2023reduced} which can catch the steady-state deformation of a droplet under stationary shear flow for high values of $\Ca$. The comparison of LT with other EMs has been instructive to qualify the importance of the main constitutive elements of the MM, $\MMCa$, and $\EMM$ models (e.g., non-linearity and non-trivial volume conservation). Our  analysis highlights that the LT model performs well only in the low capillary number regime, where deformation remains small, and it fails to capture the large deformation statistics observed at higher $\Ca$. The MM model extends the predictive range to moderate capillary numbers but still underestimates rare events and cannot reproduce the negative skewness that develops in the distribution of the normalized width ($W/R$). 
Such behavior may be attributed to the MM model’s tendency to underestimate droplet deformation, which in turn leads to an overestimation of the critical capillary number.
In contrast, both $\MMCa$ and EMM show improved agreement with FRS data across the full range of $\Ca$ explored: they reproduce the skewed and heavy-tailed distributions of all three semiaxes, including the transition from positive to negative skewness in $W/R$, associated with the emergence of cigar-like shapes near breakup~\cite{biferale2014deformation}. 

This study opens interesting perspectives for further fruitful research in the future. For example, one could extend our analysis to cases with other Reynolds numbers and/or different viscosity ratios $\lambda$, which in the present study have been kept fixed. Moreover, our study can also be considered as a starting point for the exploration of the dynamics of soft particles exhibiting additional complexities at the interface, like for example surface viscosity~\cite{barthesbiesel1985,flumerfelt1980effects} or elasticity~\cite{barthes1981time,barthes-bieselMotionDeformationElastic2016}; these interfacial properties can be introduced either in the numerical methodology that we used to get the ground truth data~\cite{guglietta2020effects,guglietta2024analytical,pelusi2023sharp} and also in the EMs~\cite{barthesbiesel1985,narsimhan2019shape,flumerfelt1980effects}. All in all, these studies could provide valuable insights to make an assessment of the validity of EMs. Beyond all the physical information that one could gather, we also remark that FRSs may become computationally expensive; hence, having the possibility to generate in shorter times ensembles of data with quantitatively validated EMs is surely welcomed for statistical analysis and/or AI studies~\cite{de2025predicting,jia2025droplet,au2023predicting,choudhury2023uncovering,tembely2022machine} on the dynamics of soft particles with complex interfaces in complex flows.

\section{Acknowledgments} 
We wish to acknowledge Fabio Bonaccorso, Luca Biferale and Maurizio Carbone for support and useful discussions. This work received funding from the European Research Council (ERC) under the European Union's Horizon 2020 research and innovation programme (grant agreement No 882340). This work was supported by the Italian Ministry of University and Research (MUR) under the FARE program (No. R2045J8XAW), project ``Smart-HEART''. MS acknowledges the support of the National Center for HPC, Big Data and Quantum Computing,  Project CN\_00000013 - CUP E83C22003230001,   Mission 4 Component 2 Investment 1.4, funded by the European Union - NextGenerationEU. Funding from Tor Vergata University project AI4HEART is also acknowledged. 

\printbibliography

\end{document}